\documentclass[onecolumn,manuscript,amsmath,amssymb,showpacs,apsrev4-1]{revtex4-1}
\usepackage{graphicx}
\usepackage{dcolumn}
\usepackage{bm}
\usepackage{SIunits}
\usepackage{verbatim}
\usepackage{placeins}

\begin{document}

\title{Coexistence of localized and itinerant magnetism in intercalated iron-selenide (Li,Fe)OHFeSe}
\author{Da-Yong Liu$^{1}$}
\email[]{dyliu@theory.issp.ac.cn}
\author{Zhe Sun$^{2}$}
\author{Liang-Jian Zou$^{1,3}$}
\email{zou@theory.issp.ac.cn}
\affiliation{1 Key Laboratory of Materials Physics, Institute of Solid State Physics, Chinese Academy of Sciences, P. O. Box 1129, Hefei, Anhui 230031, China}
\affiliation{2 National Synchrotron Radiation Laboratory, University of Science and Technology of China, Hefei, Anhui 230029, P. R. China}
\affiliation{3 Department of Physics, University of Science and Technology of China, Hefei, 230026, China}

\begin{abstract}
The electronic structure and magnetism of a new magnetic intercalation compound (Li$_{0.8}$Fe$_{0.2}$)OHFeSe are investigated theoretically. The electronic structure calculations predict that the Fe in the (Li,Fe)OH intercalated layer is in +2 valence state, ${\it i.e.}$ there is electron doping to the FeSe layer, resulting in the shrinking of the Fermi surface (FS) pocket around $\Gamma$ and a strong suppression of dynamical spin susceptibility at $M$ in comparison with the bulk FeSe compound. The ground state of the FeSe layer is a striped antiferromagnetic (SAFM) metal, while the (Li,Fe)OH layer displays a very weak localized magnetism, with an interlayer ferromagnetic (FM) coupling between the FeSe and intercalated (Li,Fe)OH layers. Moreover, the (Li,Fe)OH is more than a block layer; it is responsible for enhancing the antiferromagnetic (AFM) correlation in the FeSe layer through interlayer magnetic coupling. We propose that the magnetic spacer layer introduces a tuning mechanism for spin fluctuations associated with superconductivity in iron-based superconductors.
\end{abstract}

\vskip 300 pt

\maketitle

\section{Introduction}

Since the discovery of superconductivity in the layered iron-based compound LaO$_{1-x}$F$_{x}$FeAs \cite{JACS130-3296}, more iron pnictide and selenide compounds, including the 1111 (ROFeAs, R = La, Sm, Nd, $\it{etc.}$), 111 (AFeAs, A = Li, Na, $\it{etc.}$), 122 (BaFe$_{2}$As$_{2}$ and KFe$_{2}$Se$_{2}$, $\it{etc.}$) and 11 (FeSe and FeTe) systems \cite{Nat453-761,PRB82-180520,RMP85-849,PRB78-134514,nphy8-709,PRL102-247001}, have been successfully synthesized. Of all these materials, FeSe and its intercalated compounds A$_{x}$Fe$_{2-y}$Se$_{2}$ (A = alkali metal; K, Rb and Cs, $\it{etc.}$) have drawn extensive attention due to their interesting local interactions and Fermi surface (FS) topology which is distinct from other iron-pnictides \cite{PRL102-177003,PRL104-216405,PRL106-187001,PRL107-056401,PRB83-144511,JPCM25-125601}.
Although the superconducting transition temperature T$_{c}$ is relatively low in bulk FeSe (8 K) \cite{PNAS105-14262}, it is greatly increased by tuning the crystalline structures. For example, applying high pressure can enhance the T$_{c}$ up to 37 K in bulk FeSe \cite{PRB80-064506}, K intercalation can result in K$_{1-x}$Fe$_{2-y}$Se$_{2}$ compounds with a T$_{c}$ of 30 K \cite{PRB82-180520}, and more surprisingly, it has been found that a single layer FeSe grown on SrTiO$_{3}$ substrates can boost the T$_{c}$ \cite{Nmat12-605}. All these findings indicate the key role of the two-dimensional (2D) FeSe planes. On the other hand, due to the proximity of antiferromagnetism (AFM) and superconductivity in these iron-based compounds, magnetism has become a central issue to resolve the mystery of superconductivity in these materials. The origin of magnetism has been hotly debated and has attracted great interest about whether it comes from a FS nesting mechanism of itinerant electrons or a frustrated superexchange one of localized electrons \cite{EPL83-27006,PRL100-237003,PRL101-057003,PRL101-087004,PRL105-107004,nmat10-932,RMP85-849}.
The strong variation of T$_{c}$ with structural parameters in FeSe and its derived compounds provides a unique playground in which to investigate the nature of magnetism, the roles of localized and itinerant electrons and their connection with superconductivity in iron-based superconductors \cite{PRB84-054527,nmat8-310,PRB84-054419,nmat7-953}.

Recently, a novel iron-selenide compound (Li,Fe)OHFeSe has been found to exhibit a high superconducting transition temperature T$_{c}$ up to 40 K \cite{PRB89-020507,nmat14-325,ACIE54-293,IC54-1958}. It is isostructural with ROFeAs (R = rare earth; La, Sm, $\it{etc.}$) compounds. The anti-PbO-type layers of (Li,Fe)OH are intercalated between the anti-PbO-type FeSe layers. This leads to a very large crystal lattice parameter $c$, making the FeSe layer more 2D. There are two kinds of Fe ions in different layers, $\it{i.e.}$ Fe1 ions in the (Li,Fe)OH layer and Fe2 ions in the FeSe layer; this is significantly different from other iron-based superconductors. This case is very similar to the high-temperature cuprate superconductor YBa$_{2}$Cu$_{3}$O$_{7}$ \cite{RMP61-433}, which possesses two completely different types of Cu ions. However, the interplay of the (Li,Fe)OH and FeSe layers remains unclear; questions remain about the magnetic ground state of the (Li,Fe)OH layer and whether it is metallic or insulating. Moreover, how do the doped electrons from the (Li,Fe)OH layer affect the FS topology of the FeSe layer? We expect (Li,Fe)OHFeSe to be a very interesting model system to study how a spacer layer of magnetism interacts with the FeSe layers and how such an interaction can affect the superconductivity in FeSe layers.
%

In this paper, in order to uncover the roles of the two different kinds of Fe ions in (Li,Fe)OHFeSe, we performed first-principle calculations and disentangled the 3$d$-bands of Fe2 from other bands using Wannier functions. We find that the (Li$_{0.8}$Fe$_{0.2}$)OH layer contributes 0.2 electrons per Fe2 to the FeSe layer, forming an unique FS topology compared with the bulk FeSe compound. The magnetic ground state for Fe2 ions in the FeSe layer is a striped AFM (SAFM) state with bad metallic behavior, while the ground state of the (Li$_{0.8}$Fe$_{0.2}$)OH layer is also an SAFM state with a localized nature. Furthermore, through constructing a $J_{1}$-$J_{2}$-$J_{3}$-$J^{'}_{1}$-$J^{'}_{2}$-$J_{c}$ Heisenberg model, we find that the magnetic spacer layer introduces a tuning factor for the spin fluctuations in the FeSe layers. These results show a novel scenario different from other iron-based superconductors, which may provide new reference material to clarify the roles of nesting and magnetism. The rest of the paper is organized as follows: the calculated methods adopted are described in section 2; then the numerical results and discussions are presented in section 3; the last section is devoted to the conclusions.

\section{Method}
In our calculations, we adopt the full potential linearized augmented-plane-wave scheme based on density functional theory in the WIEN2K code \cite{WIEN2K}. Exchange and correlation effects are taken into account in the generalized gradient approximation (GGA) by Perdew, Burk, and Ernzerhof (PBE) \cite{PBE}. In order to calculate the magnetic structure, a 1 $\times$ 1 $\times$ 1 unit cell ($\sqrt{5}$ $\times$ $\sqrt{5}$ FeSe unit cell), a $\sqrt{2} \times \sqrt{2} \times 1$ supercell, and a $2 \times 2 \times 1$ supercell are used for ferromagnetic (FM) and N$\acute{e}$el AFM (NAFM), SAFM, and bi-collinear AFM (BAFM) states for the (Li$_{0.8}$Fe$_{0.2}$)OHFeSe compound, respectively. A sufficient number of $k$ points is adopted. In the disentanglement procedure, the maximally localized Wannier functions (MLWF) scheme, implemented with WANNIER90 \cite{CPC178-685} and WIEN2WANNIER \cite{CPC181-1888}, is performed.

In order to compare our numerical results with the experimental data, we adopt the structural data of (Li,Fe)OHFeSe given by neutron powder diffraction \cite{nmat14-325}. (Li,Fe)OHFeSe has a tetragonal space group $P4/nmm$ (No. 129) with lattice parameters $a$ = 3.78 ${\AA}$ ($a$ = 3.76 and 3.82 ${\AA}$ for the bulk FeSe \cite{PNAS105-14262} and FeSe/SrTiO$_{3}$ \cite{CPL31-017401}, respectively), and $c$ = 9.16 ${\AA}$ ($c$ = 5.52 ${\AA}$ for bulk FeSe), as shown in Fig.~\ref{Fig1}. Thus the (Li,Fe)OHFeSe can be regarded as bulk FeSe with a large interlayer distance. In the FeSe layer, Fe2 ions form a square lattice as in other iron-based compounds. The nearest-neighbor (N.N.) Fe2-Fe2 distance is about 2.67 $\AA$, and the next nearest-neighbor (N.N.N.) distance is about $\sqrt{2}$ times that of the N.N. distance, 3.78 $\AA$. One Fe2 atom is surrounded by four Se atoms which form a tetrahedral environment. Note that in the (Li$_{0.8}$Fe$_{0.2}$)OH layer, the Li and Fe atoms are randomly distributed on a square lattice with a ratio of 4:1. In order to construct a supercell, we assumed Li and Fe1 atoms in the (Li$_{0.8}$Fe$_{0.2}$)OH layer to be orderly distributed. (Li$_{0.8}$Fe$_{0.2}$)OHFeSe has a $\sqrt{5}$ $\times$ $\sqrt{5}$ supercell compared with a conventional FeSe unit cell. In the (Li$_{0.8}$Fe$_{0.2}$)OH layer, Fe1 ions also form a $\sqrt{5}$ $\times$ $\sqrt{5}$ square lattice comparable with a Fe2 lattice in the FeSe layer. The N.N. and N.N.N. Fe1-Fe1 distance is 5.97 $\AA$ and 8.45 $\AA$, respectively. Notice that the distance of Fe1-O in the (Li$_{0.8}$Fe$_{0.2}$)OH layer is 2.01 $\AA$, smaller than that of Fe2-Se (2.40 $\AA$) in the FeSe layer.

\section{Results and discussions}

\subsection{Nonmagnetic (NM) state}
\begin{figure}[htbp]
\hspace*{-2mm}
\centering
\includegraphics[trim = 0mm 0mm 0mm 0mm, clip=true, width=0.7 \columnwidth]{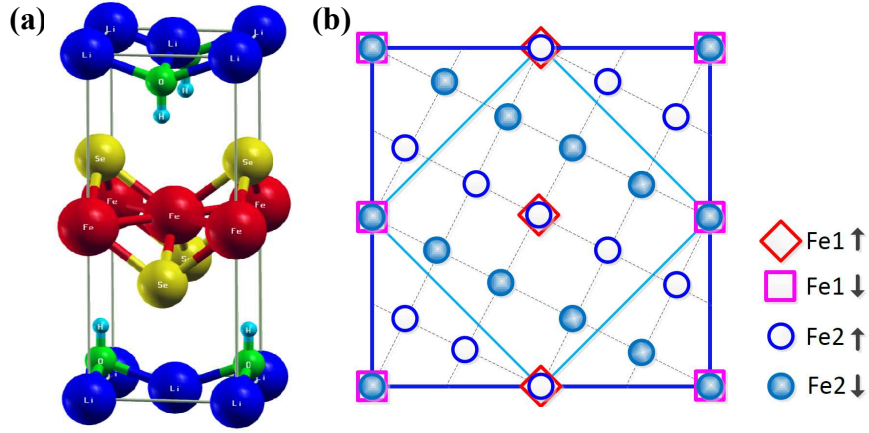}
\caption{(Color online) (a) Crystal structure of (Li,Fe)OHFeSe with Li, Fe, O and Se atoms in blue, red, green, and yellow, respectively, and (b) magnetic structure of (Li$_{0.8}$Fe$_{0.2}$)OHFeSe with an SAFM(FeSe)-FM(interlayer)-SAFM((Li$_{0.8}$Fe$_{0.2}$)OH) configuration.}
\label{Fig1}
\end{figure}

\begin{figure}[htbp]\centering
\includegraphics[angle=0, width=0.7 \columnwidth]{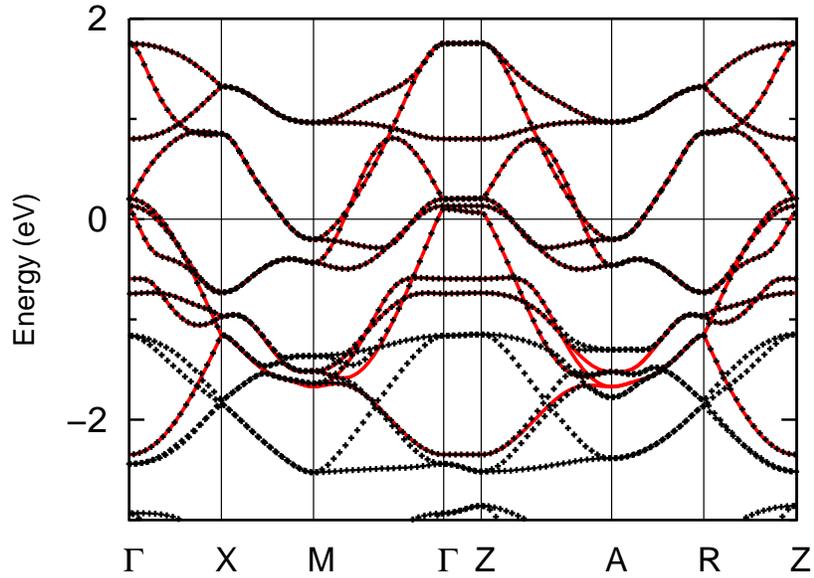}
\caption{(Color online) Band structures of NM LiOHFeSe. The plus symbols represent the band structure obtained by the GGA, and the solid lines show the band structures projected on the Fe2 3$d$ orbitals using the MLWF method.}
\label{Fig2}
\end{figure}

\begin{figure}[htbp]\centering
\includegraphics[angle=0, width=0.85 \columnwidth]{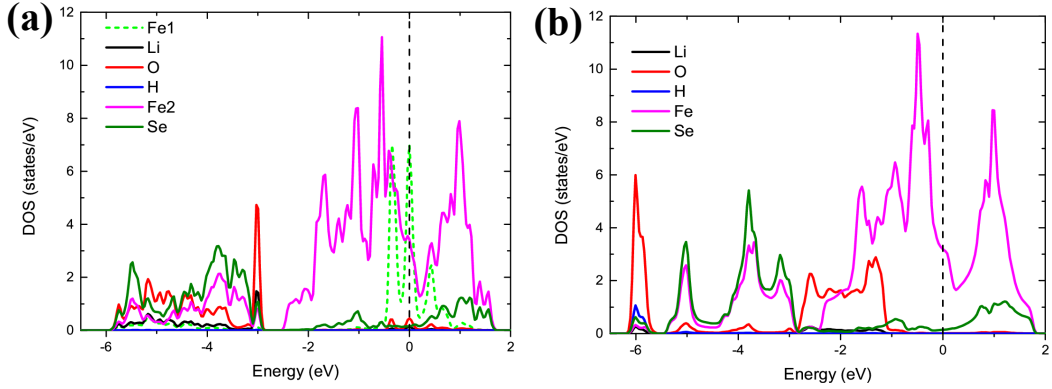}
\caption{(Color online) Atomic-resolved DOS of NM (Li$_{0.8}$Fe$_{0.2}$)OHFeSe (a) and LiOHFeSe (b) compounds.}
\label{Fig3}
\end{figure}

We first study the NM phase of an ideal LiOHFeSe compound \cite{PLA379-2106}, which can serve as a reference for the (Li$_{0.8}$Fe$_{0.2}$)OHFeSe compound. The band structure of the NM state in LiOHFeSe within the GGA is shown in Fig.~\ref{Fig2}.
The bands involved with 3$d$ orbitals of Fe2 are disentangled from other bands using the MLWF method and a ten-band tight-binding model is constructed. For comparison, the atomic-resolved density of states (DOS) of NM (Li$_{0.8}$Fe$_{0.2}$)OHFeSe and LiOHFeSe are given in Fig.~\ref{Fig3}. Due to the large $c$-axis lattice parameter, the FeSe layer is nearly separated by the (Li,Fe)OH spacer layer. Thus the electronic structures can be seen as a composite one, which is contributed from both Fe1 and Fe2, respectively. It is noticed that the DOS of Fe2 in the FeSe layer in both the realistic (Li$_{0.8}$Fe$_{0.2}$)OHFeSe and ideal LiOHFeSe compounds are nearly the same. This implies the spacer layer (Li$_{0.8}$Fe$_{0.2}$)OH is isolate. Thus we can only investigate the electronic properties of the FeSe layer of the LiOHFeSe compound separately.
%
According to the analysis of the Fe1 3$d$ orbital occupation, we find that Fe1 ion in the (Li,Fe)OH layer possesses +2 valence with six electrons occupied, which is consistent with the recent M$\ddot{o}$ssbauer experiment \cite{JAC652-470}, indicating that the FeSe layer is electron doped by the (Li$_{0.8}$Fe$_{0.2}$)OH layer.

\begin{figure}[htbp]\centering
\includegraphics[trim = 0mm 0mm 0mm 0mm, clip=true, width=0.7 \columnwidth]{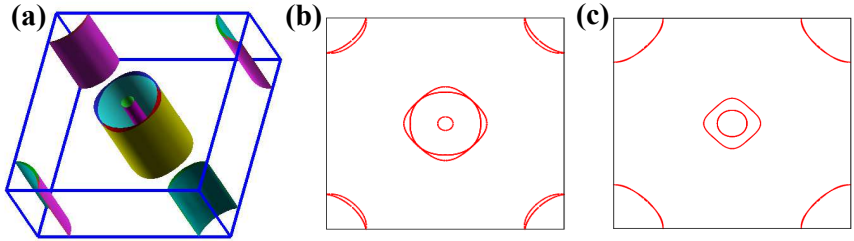}
\caption{(Color online) (a) Fermi surface, and (b) its $k_{z}=0$ 2D cut plot of LiOHFeSe obtained by the GGA, and (c) 2D cut for 0.2 electrons doping per Fe site corresponding to the (Li$_{0.8}$Fe$_{0.2}$)OHFeSe compound.
}\label{Fig4}
\end{figure}

The unique FS topologies in LiOHFeSe and its $k_{z}=0$ 2D cut plot, as shown in Fig.~\ref{Fig4}(a) and (b), are very different from other iron-based superconductors \cite{PRB84-064435,physicab407-1139,JPCM25-125601}, consisting of a small hole cylinder and two inner hole cylinders centered at $\Gamma$ and two electron cylinders at $M$ arising from Fe2 3$d$ bands.
As mentioned above, (Li$_{0.8}$Fe$_{0.2}$)OHFeSe can be regarded as a LiOHFeSe system doped by 0.2 electrons per Fe2 ions. In fact the angle resolved photoemission spectroscopy (ARPES) experiment \cite{ncomms7-10608} also supports the electron doping from Fe1 ions. The 2D cut FS of the doped LiOHFeSe by 0.2 electrons per Fe ions is plotted in Fig.~\ref{Fig4}(c). It is clearly found that the hole cylinders at $\Gamma$ shrink, while the nearly degenerate electron cylinders at $M$ enlarge. The (Li,Fe)OH layer therefore leads to a sharp change in the FS topology of the FeSe layer, which is very different from that of the bulk FeSe compound.

\begin{figure}[htbp]\centering
\includegraphics[trim = 0mm 0mm 0mm 0mm, clip=true, width=0.7 \columnwidth]{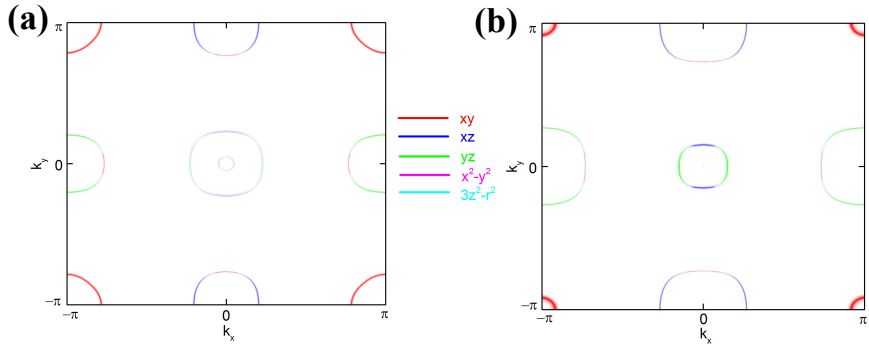}
\caption{(Color online) The unfolded orbital-resolved FS for (a) LiOHFeSe and (b) (Li$_{0.8}$Fe$_{0.2}$)OHFeSe in one-Fe BZ.
}\label{Fig5}
\end{figure}

\begin{figure}[htbp]\centering
\includegraphics[trim = 0mm 0mm 0mm 0mm, clip=true, width=0.7 \columnwidth]{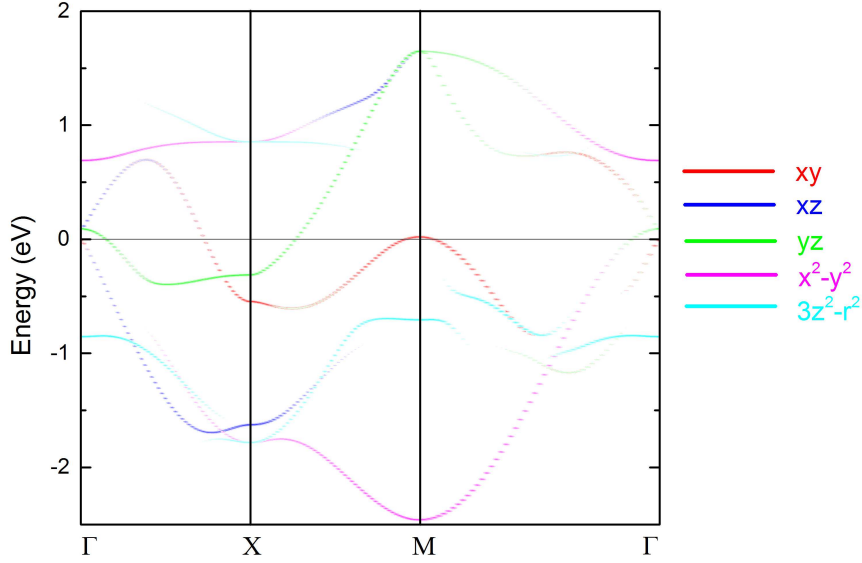}
\caption{(Color online) The unfolded orbital-resolved band structures of (Li$_{0.8}$Fe$_{0.2}$)OHFeSe in one-Fe BZ.
}\label{Fig6}
\end{figure}

Considering the fact that the two inequivalent Fe ions in the doped LiOHFeSe system lead to a folded electronic structure, we also unfold it to one Fe scenario based on the band unfolding technique \cite{PRL104-216401}. The unfolded orbital-resolved FS and band structure corresponding to one-Fe Brillouin zone (BZ) are shown in Fig.~\ref{Fig5} and Fig.~\ref{Fig6}, respectively.
The FS topology of (Li$_{0.8}$Fe$_{0.2}$)OHFeSe, as shown in Fig.~\ref{Fig5}(b), is greatly consistent with the result observed by the ARPES experiment \cite{ncomms7-10608}. The small hole FS mainly contributes from the $xy$ orbital, while the large hole FS contributes from the $xz$ and $yz$ orbitals, and the two degenerate electron FSs from both the $xy$ and $xz$, $yz$ orbitals.

As seen in Fig.~\ref{Fig6}, the unfolded band structures are very similar to those of the FeSe compound \cite{PRB78-134514}; in addition, a Fermi level shifts upwards by 0.1 eV. This indicates that the (Li$_{0.8}$Fe$_{0.2}$)OH spacer layer only provides electron doping and chemical pressure (strain) with respective to the FeSe system. The unfolded band structure displays an almost one-Fe (five-orbital) picture, which is also observed by ARPES experiments in the iron-based compounds CsFe$_{2}$As$_{2}$ and RbFe$_{2}$As$_{2}$ \cite{PRB92-184512}.

\subsection{Magnetic instability}
\begin{figure}[htbp]\centering
\includegraphics[trim = 0mm 0mm 0mm 0mm, clip=true, width=0.8 \columnwidth]{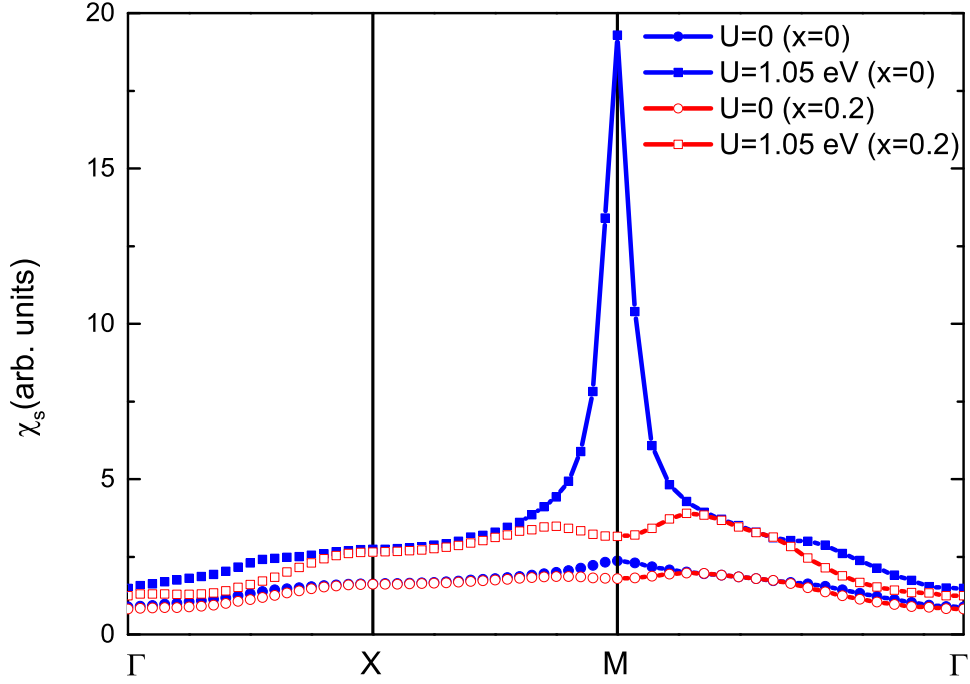}
\caption{(Color online) Bare and RPA spin susceptibilities along the high symmetry path in (Li$_{1-x}$Fe$_{x}$)OHFeSe at $x$=0 and $x$=0.2, respectively.
}\label{Fig7}
\end{figure}

In order to explore whether there is a FS nesting in (Li,Fe)OHFeSe and its associated magnetic instability, we calculate the bare spin susceptibility given by \cite{physicab407-1139,NJP11-025016}
\begin{eqnarray}
  \chi_{0}(q)=\frac{1}{N}\sum_{\substack{\vec{k},n,m}}
  \frac{f(\varepsilon_{n}(\vec{k}))-f(\varepsilon_{m}(\vec{k}+\vec{q}))}
  {\varepsilon_{m}(\vec{k}+\vec{q})-\varepsilon_{m}(\vec{k})+i\eta}
\label{Eq1}
\end{eqnarray}
Considering the moderate electronic correlation in iron-based systems, the multi-orbital random phase approximation (RPA) spin susceptibility $\chi_{s}^{RPA}$ is written as
\begin{eqnarray}
  \chi_{s}^{RPA}(\mathbf{q}, i\omega) &=& \chi_{0}(\mathbf{q},i\omega)[\mathbb{I}-\Gamma_{s}\chi_{0}(\mathbf{q},i\omega)]^{-1}
\label{Eq2}
\end{eqnarray}
where $\chi_{0}$ is the bare susceptibility defined in Eq.(~\ref{Eq1}), and the nonzero components of the interaction matrices $\Gamma_{s}$ are given as $(\Gamma_{s})_{aa,aa}=U$, $(\Gamma_{s})_{aa,bb}=J_{H}$, $(\Gamma_{s})_{ab,ab}=U'$, and $(\Gamma_{s})_{ab,ba}=J_{H}$ with the orbital indices $a\neq b$.
The obtained dynamical spin susceptibilities based on the ten-band model are plotted in Fig.~\ref{Fig7}, where the spin susceptibilities of both the $x$=0 and $x$=0.2 cases are given for comparison.
It clearly shows a sharp peak at $M$ with $Q \sim (\pi,\pi)$ at $x$=0 in the LiOHFeSe system, favoring an SAFM configuration. This suggests good nesting is driven by the Fe2 3$d$ orbitals of the FeSe layer, similar to other iron-based superconductors \cite{physicab407-1139,PRB93-195148}. In contrast, for the $x$=0.2 case, the lack of sharp peaks in the susceptibilities indicates poor FS nesting in (Li$_{0.8}$Fe$_{0.2}$)OHFeSe. This suggests that the electron doping in (Li$_{0.8}$Fe$_{0.2}$)OHFeSe does not favor the SAFM long-range ordering state.

\subsection{Magnetic ground state}
To further confirm the magnetic ground state of (Li$_{0.8}$Fe$_{0.2}$)OHFeSe, several magnetic structures are investigated to compare their energies, including the NM, FM, NAFM, SAFM and BAFM states both in the FeSe layer and the (Li,Fe)OH layer, with interlayer FM or AFM couplings. We find that the intralayer SAFM configurations both in the FeSe layer and in the (Li,Fe)OH layer are the lowest, though various interlayer magnetic couplings are very close to each other. The lowest state is the SAFM phase with weak interlayer FM coupling, as shown in Fig.~\ref{Fig1}(b). The calculated magnetic moments are 3.5 $\mu_{B}$ for Fe1 and 2.4 $\mu_{B}$ for Fe2, respectively. The comparison of various magnetic configurations is listed in Table~\ref{Table1}.

\begin{figure}[htbp]\centering
\includegraphics[angle=0, width=0.7 \columnwidth]{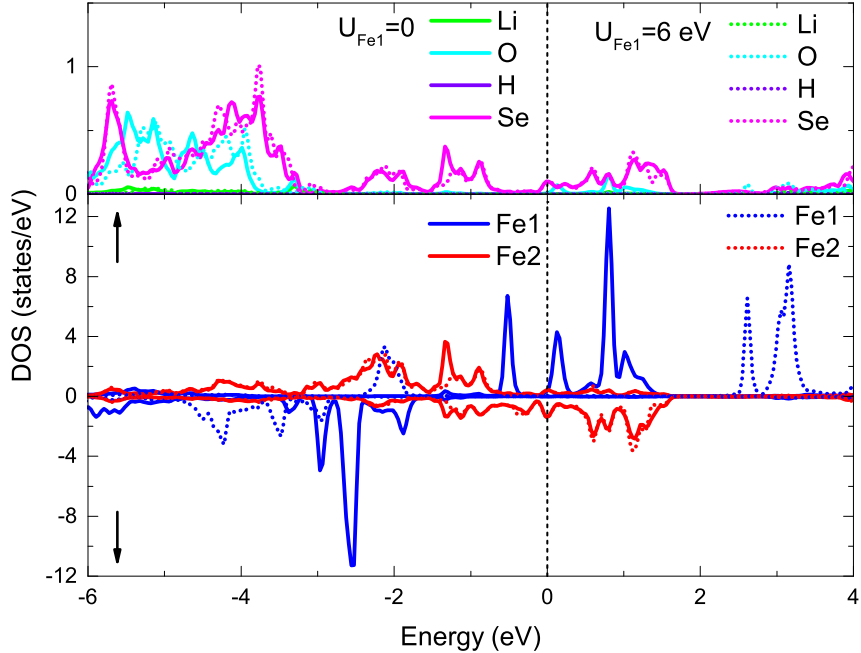}
\caption{(Color online) Atomic-resolved DOS of the SAFM state in (Li$_{0.8}$Fe$_{0.2}$)OHFeSe by the GGA (solid line) and the GGA+$U$ (dot line) with $U_{Fe1}$ = 6 eV.}
\label{Fig8}
\end{figure}

In fact, although many experimental and theoretical studies have been performed, the electronic and magnetic properties of the (Li,Fe)OH layer are still under debate. In Fig.~\ref{Fig8}, the atomic-resolved DOS of the magnetic ground state with the SAFM-FM-SAFM configuration is plotted. It is clearly seen that the states contributed by the Fe1 ions at the Fermi level are very small. Further, considering the Coulomb electronic correlation, these states would become fully gaped within the GGA+$U$ framework, leading to a localized magnetism. This implies that the electronic correlations of both the Fe1 and Fe2 ions should perhaps be considered in the (Li,Fe)OHFeSe compound.
Here we only consider the case of strong electronic correlation in the (Li,Fe)OH layer. Our results
show that for $U_{Fe1}$ $\gtrsim$ 2 eV, the state of Fe1 would be insulated. Considering that the Coulomb interaction parameter of iron is generally smaller than 7 eV, we choose $U_{Fe1}$ = 6 eV; the corresponding GGA+$U$ result is also shown in Fig.~\ref{Fig8}.
On the other hand, the Fe2$^{2+}$ (3$d^{6}$) ions display bad metallic behavior with itinerant magnetism. Thus there is a coexistence of localized and itinerant magnetism in (Li,Fe)OHFeSe, similar to SmFeAsO$_{1-x}$F$_{x}$ \cite{Nat453-761}.
For comparison, the relative energy difference per Fe ions of different magnetic configurations with respective to the total energy of the NM state in (Li$_{0.8}$Fe$_{0.2}$)OHFeSe is listed in Table~\ref{Table1}. Note that the magnetic configurations are denoted by such a form with Fe1/Fe1-Fe1/Fe2-Fe2/Fe2 type, denoting the intralayer coupling between Fe1 ions in the (Li$_{0.8}$Fe$_{0.2}$)OH layer, the interlayer coupling between Fe1 and Fe2, and the intralayer coupling between the Fe2 ions in the FeSe layer, respectively.
\begin{table}[htpb]
\caption{Relative energy difference per Fe2 ions of different magnetic configurations with respect to the NM phase in (Li$_{0.8}$Fe$_{0.2}$)OHFeSe. The magnetic moments of Fe1 and Fe2 are also given.}
\begin{center}
\begin{tabular}{lrcccccc}
\hline\hline
\hline
& Magnetic configuration       & $\Delta{E}$/Fe2 (meV) &   Magnetic moment & ($\mu_{B}$)   \\
\hline
&Fe1/Fe1-Fe1/Fe2-Fe2/Fe2       &$\Delta{(E-E(NM))}$/Fe2  &Fe1     &Fe2                \\
\hline
\hline
& NM                           &0                   &0       &0                 \\
& FM-AFM-FM                     &-413.37             &3.46    &1.99               \\
& FM-AFM/FM-NAFM                 &-603.57             &3.4     &1.94             \\
& FM-AFM/FM-SAFM                 &-638.80             &3.4     &2.2              \\
& NAFM-AFM/FM-SAFM               &-638.83             &3.4     &2.2             \\
& SAFM-AFM-SAFM                 &-638.77             &3.4     &2.2               \\
& SAFM-FM-SAFM                  &-638.98             &3.4     &2.2              \\
& BAFM-FM-BAFM                  &-579.67             &3.4     &2.2              \\
\hline
\hline\hline
\end{tabular}
\end{center}
\label{Table1}
\end{table}
We note that the magnetic state with the same spin configurations both for Fe1 in the (Li,Fe)OH layer and for Fe2 in the FeSe layer is more stable than others, indicating that the magnetic state of the (Li,Fe)OH layer is determined by that of the FeSe layer due to strong interlayer magnetic coupling in comparison with the planar one in the (Li,Fe)OH layer.

In order to describe the magnetic interactions between spins in this system, a $J_{1}$-$J_{2}$-$J_{3}$-$J^{'}_{1}$-$J^{'}_{2}$-$J_{c}$ Heisenberg model is constructed as follows,
\begin{equation}
\label{eq.1}
\begin{aligned}
H=&J_{1}\sum_{\substack{<ij>_{Fe2}}}\vec{s}_{i}\cdot\vec{s}_{j}
+J_{2}\sum_{\substack{<<ij>>_{Fe2}}}\vec{s}_{i}\cdot\vec{s}_{j}\\
\nonumber
&+J_{3}\sum_{\substack{<<<ij>>>_{Fe2}}}\vec{s}_{i}\cdot\vec{s}_{j}
+J^{'}_{1}\sum_{\substack{<mn>_{Fe1}}}\vec{S}_{m}\cdot\vec{S}_{n}\\
\nonumber
&+J^{'}_{2}\sum_{\substack{<<mn>>_{Fe1}}}\vec{S}_{m}\cdot\vec{S}_{n}
+J_{c}\sum_{\substack{<im>_{c}}}\vec{S}_{m}\cdot\vec{s}_{i},
\end{aligned}
\end{equation}
where $J_{1}$, $J_{2}$ and $J_{3}$ are the N.N., N.N.N. and third-N.N. intralayer magnetic couplings in the FeSe layer with spin $s$ of Fe2 ions, respectively; $J^{'}_{1}$ and $J^{'}_{2}$ are the N.N. and N.N.N. intralayer magnetic coupling strengths in the (Li$_{0.8}$Fe$_{0.2}$)OH layer with spin $S$ of Fe1 ions, and $J_{c}$ is the N.N. interlayer magnetic coupling strength between Fe1 and Fe2 ions. The spin exchange parameters can be obtained by the differences of total energies between different magnetic structures, which are listed in the following,
\begin{equation}
\label{eq.2}
\begin{aligned}
\Delta E(NAFM-FM)_{Fe_{2}}&=&&-4J_{1}s^{2} \\
\nonumber
\Delta E(NAFM-SAFM)_{Fe_{2}}&=&&-2(J_{1}-2J_{2})s^{2} \\
\nonumber
\Delta E(NAFM-BAFM)_{Fe_{2}}&=&&-2(J_{1}-J_{2}-2J_{3})s^{2} \\
\nonumber
\Delta E(NAFM-FM)_{Fe_{1}}&=&&-4J^{'}_{1}S^{2} \\
\nonumber
\Delta E(NAFM-SAFM)_{Fe_{1}}&=&&-2(J^{'}_{1}-2J^{'}_{2})S^{2} \\
\nonumber
\Delta E(AFM-FM)_{Fe_{1,2}}&=&&-J_{c}Ss
\end{aligned}
\end{equation}
According to Table~\ref{Table1}, the calculated spin exchange parameters are $J_{1}$ = 47.5 meV/$s^{2}$,
$J_{2}$ = 32.6 meV/$s^{2}$, $J_{3}$ = 1.75 meV/$s^{2}$, $J^{'}_{1}$ = 0.1 meV/$S^{2}$,
$J^{'}_{2}$ = 0.11 meV/$S^{2}$, and $J_{c}$ = $-$0.54 meV/$Ss$, respectively. Except for the interlayer FM coupling, the positive values of all the spin exchange parameters favor an AFM configuration. Moreover, the relationship $J_{2}>J_{1}/2$ can result in a strong magnetic frustration in the FeSe layer of (Li$_{0.8}$Fe$_{0.2}$)OHFeSe, similar as in LaOFeAs. Consequently the magnetic structure can be well understood within this Heisenberg model.

Moreover, we have also investigated the influence of Fe1 magnetic ions in the (Li$_{0.8}$Fe$_{0.2}$)OH layer on the FeSe layer. We find that the spin configuration of Fe2 ions pins that of Fe1 ions through the interlayer magnetic coupling between the Fe1 and Fe2 ions. As a consequence, the magnetic state of the (Li$_{0.8}$Fe$_{0.2}$)OH layer would affect the magnetic state or superconducting pairing of the FeSe layers. Note that it is known that the N.N.N. exchange coupling plays a key role in the superconductivity of iron-based superconductors \cite{PRL101-057003}. Thus the strong magnetic fluctuation induced by the (Li$_{0.8}$Fe$_{0.2}$)OH layer is possibly responsible for the high superconducting transition temperature T$_{c}$ in (Li,Fe)OHFeSe. Therefore our results demonstrate that the interplay between the different magnetic layers is a tuning factor for spin fluctuations associated with superconductivity in iron-based superconductors. In fact, the enhancement of magnetism is also observed experimentally in other iron-based materials, which is usually ascribed to the influence of the interstitial Fe \cite{PRL107-216403,PRL108-107002}.

It is worth noting that there are many sensitive factors which may seriously affect the magnetism in (Li$_{0.8}$Fe$_{0.2}$)OHFeSe. One is the possible stoichiometric problem extensively existing in iron-based superconductors; for instance, excess Fe in the experiments may lead to contradictory results between theory and experiment, $\it{i.e.}$ theoretically-predicted magnetic order was not observed experimentally for FeSe \cite{PRL102-177003,nmat8-630,PRB78-174502} and LiFeAs \cite{PRB78-094511,PRB81-140511}. Another issue is the cation-disorder of Li and Fe ions in the (Li,Fe)OH layer of the (Li$_{0.8}$Fe$_{0.2}$)OHFeSe compound.
Thus further neutron scattering and nuclear magnetic resonance experiments are needed to investigate the magnetism in (Li,Fe)OHFeSe.

\section{Conclusions}
In summary, we have performed first-principle calculations and theoretical analysis to the electronic structures and magnetic properties of (Li,Fe)OHFeSe compounds. We find that the low energy physics of this novel iron-selenide superconductor (Li,Fe)OHFeSe is dominated by the FeSe layer, while the (Li,Fe)OH layer is not only a Mott insulating block layer, but is also responsible for the enhancement of the N.N.N. AFM correlations in the FeSe layer. We find that the ground state is SAFM configurations in both the FeSe and (Li,Fe)OH layers with weak FM interlayer coupling and, furthermore, the magnetic coupling strengths of Fe spins are evaluated. The coexistence of the localized and itinerant magnetism in (Li,Fe)OHFeSe provides a good example for investigating the interplay of localized and itinerant electrons, and the interplay of the magnetism and superconductivity.
\\

\noindent
\textbf{Acknowledgments}\\
\noindent
We acknowledge that X-H Chen provided us with the early experimental data. This work was supported by the National Sciences Foundation of China under Grant Nos. 11574315, 11274310, 11474287 and 11190022. Numerical calculations were performed at the Center for Computational Science of CASHIPS, the ScGrid of Supercomputing Center and Computer Net work Information Center of Chinese Academy of Science. The crystal structure and FS were visualised with XCrysDen \cite{xcrys}.


\section*{References}
\begin{thebibliography}{10}

\bibitem{JACS130-3296}
Kamihara Y, Watanabe T, Hirano M and Hosono H 2008
{\it J. Am. Chem. Soc.} {\bf 130} 3296

\bibitem{Nat453-761}
Chen X H, Wu T, Wu G, Liu R H, Chen H and Fang D F 2008
{\it Nature} {\bf 453} 761

\bibitem{PRB82-180520}
Guo J G, Jin S F, Wang G, Wang S C, Zhu K X, Zhou T T, He M and Chen X L 2010
{\it Phys. Rev. B} {\bf 82} 180520(R)

\bibitem{RMP85-849}
Dagotto E 2013
{\it Rev. Mod. Phys.} {\bf 85} 849

\bibitem{PRB78-134514}
Subedi A, Zhang L J, Singh D J and Du M H 2008
{\it Phys. Rev. B} {\bf 78} 134514

\bibitem{nphy8-709}
Dai P C, Hu J P and Dagotto E 2012
{\it Nature Phys.} {\bf 8} 709

\bibitem{PRL102-247001}
Bao W, Qiu Y, Huang Q, Green M A, Zajdel P, Fitzsimmons M R, Zhernenkov M, Chang S, Fang M H, Qian B, Vehstedt E K, Yang J H, Pham H M, Spinu L and Mao Z Q 2009
{\it Phys. Rev. Lett.} {\bf 102} 247001

\bibitem{PRL102-177003}
Ma F J, Ji W, Hu J P, Lu Z Y and Xiang T 2009
{\it Phys. Rev. Lett.} {\bf 102} 177003

\bibitem{PRL104-216405}
Zhu J X, Yu R, Wang H D, Zhao L L, Jones M D, Dai J H, Abrahams E, Morosan E, Fang M H and Si Q M 2010
{\it Phys. Rev. Lett.} {\bf 104} 216405

\bibitem{PRL106-187001}
Qian T, Wang X P, Jin W C, Zhang P, Richard P, Xu G, Dai X, Fang Z, Guo J G, Chen X L and Ding H 2011
{\it Phys. Rev. Lett.} {\bf 106} 187001

\bibitem{PRL107-056401}
Cao C and Dai J H 2011
{\it Phys. Rev. Lett.} {\bf 107} 056401

\bibitem{PRB83-144511}
Zeng B, Shen B, Chen G F, He J B, Wang D M, Li C H and Wen H H 2011
{\it Phys. Rev. B} {\bf 83} 144511

\bibitem{JPCM25-125601}
Liu D Y, Quan Y M, Zheng X J, Yu X L and Zou L J 2013
{\it J. Phys.: Condens. Matter} {\bf 25} 125601

\bibitem{PNAS105-14262}
Hsu F C, Luo J Y, Yeh K W, Chen T K, Huang T W, Wu P M, Lee Y C, Huang Y L, Chu Y Y, Yan D C and Wu M K 2008
{\it Proc. Natl. Acad. Sci. U.S.A.} {\bf 105} 14262

\bibitem{PRB80-064506}
Margadonna S, Takabayashi Y, Ohishi Y, Mizuguchi Y, Takano Y, Kagayama T, Nakagawa T, Takata M and Prassides K 2009
{\it Phys. Rev. B} {\bf 80} 064506

\bibitem{Nmat12-605}
He S L, He J F, Zhang W H, Zhao L, Liu D F, Liu X, Mou D X,	Ou Y B, Wang Q Y, Li Z, Wang L L, Peng Y Y, Liu Y, Chen C Y, Yu L, Liu G D, Dong X L, Zhang J, Chen C T, Xu Z Y, Chen X, Ma X C, Xue Q K and Zhou X J 2013
{\it Nature Mat.} {\bf 12} 605

\bibitem{EPL83-27006}
Dong J, Zhang H J, Xu G, Li Z, Li G, Hu W Z, Wu D, Chen G F, Dai X, Luo J L, Fang Z and Wang N L 2008
{\it Europhys. Lett.} {\bf 83} 27006

\bibitem{PRL100-237003}
Singh D J and Du M H 2008
{\it Phys. Rev. Lett.} {\bf 100} 237003

\bibitem{PRL101-057003}
Mazin I I, Singh D J, Johannes M D and Du M H 2008
{\it Phys. Rev. Lett.} {\bf 101} 057003

\bibitem{PRL101-087004}
Kuroki K, Onari S, Arita R, Usui H, Tanaka Y, Kontani H and Aoki H 2008
{\it Phys. Rev. Lett.} {\bf 101} 087004

\bibitem{PRL105-107004}
Yin W G, Lee C C and Ku W 2010
{\it Phys. Rev. Lett.} {\bf 105} 107004

\bibitem{nmat10-932}
Yin Z P, Haule K and Kotliar G 2011
{\it Nature Mat.} {\bf 10} 932

\bibitem{PRB84-054527}
You Y Z, Yang F, Kou S P and Weng Z Y 2011
{\it Phys. Rev. B} {\bf 84} 054527

\bibitem{nmat8-310}
Drew A J, Niedermayer Ch, Baker P J, Pratt F L, Blundell S J, Lancaster T, Liu R H, Wu G, Chen X H, Watanabe I, Malik V K, Dubroka A, R$\ddot{o}$ssle M, Kim K W, Baines C and Bernhard C 2009
{\it Nature Mat.} {\bf 8} 310

\bibitem{PRB84-054419}
Nandi S, Su Y, Xiao Y, Price S, Wang X F, Chen X H, Herrero-Mart$\acute{\i}$n J, Mazzoli C, Walker H C, Paolasini L, Francoual S, Shukla D K, Strempfer J, Chatterji T, Kumar C M N, Mittal R, R${\o}$nnow H M, R$\ddot{u}$egg Ch, McMorrow D F and Br$\ddot{u}$ckel Th 2011
{\it Phys. Rev. B} {\bf 84} 054419

\bibitem{nmat7-953}
Zhao J, Huang Q, Cruz C De La, Li S L, Lynn J W, Chen Y, Green M A, Chen G F, Li G, Li Z, Luo J L, Wang N L and Dai P C 2008
{\it Nature Mat.} {\bf 7} 953

\bibitem{PRB89-020507}
Lu X F, Wang N Z, Zhang G H, Luo X G, Ma Z M, Lei B, Huang F Q and Chen X H 2014
{\it Phys. Rev. B} {\bf 89} 020507(R)

\bibitem{nmat14-325}
Lu X F, Wang N Z, Wu H, Wu Y P, Zhao D, Zeng X Z, Luo X G, Wu T, Bao W,
Zhang G H, Huang F Q, Huang Q Z and Chen X H 2015
{\it Nature Mat.} {\bf 14} 325

\bibitem{ACIE54-293}
Pachmayr U, Nitsche F, Luetkens H, Kamusella S, Br$\ddot{u}$ckner F, Sarkar R, Klauss H H, and Johrendt D 2015
{\it Angew. Chem. Int. Ed.} {\bf 54} 293

\bibitem{IC54-1958}
Sun H L, Woodruff D N, Cassidy S J, Allcroft G M, Sedlmaier S J, Thompson A L, Bingham P A, Forder S D, Cartenet S, Mary N, Ramos S, Foronda F R, Williams B H, Li X D, Blundell S J, and Clarke S J 2015
{\it Inorg. Chem.} {\bf 54} 1958

\bibitem{RMP61-433}
Pickett W E 1989
{\it Rev. Mod. Phys.} {\bf 61} 433

\bibitem{WIEN2K}
Blaha P, Schwarz K, Madsen G, Kvasnicka D and Luitz J 2001:
"Computer Code WIEN2k, an augmented plane wave plus local orbitals
program for calculating crystal properties", Karlheinz
Schwarz, Technische Universit$\ddot{s}$t Wien, Austria

\bibitem{PBE}
Perdew J P, Burke K and Ernzerhof M 1996
{\it Phys. Rev. Lett.} {\bf 77} 3865

\bibitem{CPC178-685}
Mostofi A A, Yates J R, Lee Y S, Souza I, Vanderbilt D and Marzari N 2008
{\it Comput. Phys. Commun.} {\bf 178} 685

\bibitem{CPC181-1888}
Kunes J, Arita R, Wissgott P, Toschi A, Ikeda H and Held K 2010
{\it Comput. Phys. Commun.} {\bf 181} 1888

\bibitem{CPL31-017401}
Zhang W H, Sun Y, Zhang J S, Li F S, Guo M H, Zhao Y F, Zhang H M, Peng J P, Xing Y, Wang H C, Fujita T, Hirata A, Li Z, Ding H, Tang C J, Wang M, Wang Q Y, He K, Ji S H, Chen X, Wang J F, Xia Z C, Li L, Wang Y Y, Wang J and Wang L L 2014
{\it Chin. Phys. Lett.} {\bf 31} 017401

\bibitem{PLA379-2106}
Wang G T, Yi X, and Shi X B  2015
{\it Phys. Lett. A} {\bf 379} 2106

\bibitem{JAC652-470}
Nejadsattari F and Stadnik Z M 2015
{\it J. Alloys Compd.} {\bf 652} 470

\bibitem{PRB84-064435}
Liu D Y, Quan Y M, Chen D M, Zou L J and Lin H Q 2011
{\it Phys. Rev. B} {\bf 84} 064435

\bibitem{physicab407-1139}
Liu D Y, Quan Y M, Zeng Z and Zou L J 2012
{\it Physica B} {\bf 407} 1139

\bibitem{ncomms7-10608}
Zhao L, Liang A J, Yuan D N, Hu Y, Liu D F, Huang J W, He S L,
Shen B, Xu Y, Liu X, Yu L, Liu G D, Zhou H X, Huang Y L, Dong X L,
Zhou F, Liu K, Lu Z Y, Zhao Z X, Chen C T, Xu Z A and Zhou X J 2015
{\it  Nat. Commun.} {\bf 7} 10608

\bibitem{PRL104-216401}
Ku W, Berlijn T and Lee C C 2010
{\it Phys. Rev. Lett.} {\bf 104} 216401

\bibitem{PRB92-184512}
Kong S, Liu D Y, Cui S T, Ju S L, Wang A F, Luo X G, Zou L J, Chen X H, Zhang G B and Sun Z 2015
{\it Phys. Rev. B} {\bf 92} 184512

\bibitem{NJP11-025016}
Graser S, Maier T A, Hirschfeld P J and Scalapino D J 2009
{\it New J. Phys.} {\bf 11} 025016

\bibitem{PRB93-195148}
Huang Y N, Liu D Y, Zou L J and Pickett W E 2016
{\it Phys. Rev. B} {\bf 93} 195148

\bibitem{PRL107-216403}
Zaliznyak I A, Xu Z J, Tranquada J M, Gu G D, Tsvelik A M and Stone M B 2011
{\it Phys. Rev. Lett.} {\bf 107} 216403

\bibitem{PRL108-107002}
Thampy V, Kang J, Rodriguez-Rivera J A, Bao W, Savici A T, Hu J, Liu T J, Qian B, Fobes D,
Mao Z Q, Fu C B, Chen W C, Ye Q, Erwin R W, Gentile T R, Tesanovic Z and Broholm C 2012
{\it Phys. Rev. Lett.} {\bf 108} 107002

\bibitem{nmat8-630}
Medvedev S, McQueen T M, Troyan I A, Palasyuk T, Eremets M I, Cava R J, Naghavi S, Casper F, Ksenofontov V, Wortmann G and Felser C 2009
{\it Nature Mat.} {\bf 8} 630

\bibitem{PRB78-174502}
Lee K W, Pardo V and Pickett W E 2008
{\it Phys. Rev. B} {\bf 78} 174502

\bibitem{PRB78-094511}
Singh D J 2008
{\it Phys. Rev. B} {\bf 78} 094511

\bibitem{PRB81-140511}
Jegli$\check{c}$ P, Poto$\check{c}$nik A, Klanj$\check{s}$ek M, Bobnar M, Jagodi$\check{c}$ M, Koch K, Rosner H, Margadonna S, Lv B, Guloy A M and Ar$\check{c}$on D 2010
{\it Phys. Rev. B} {\bf 81} 140511(R)

\bibitem{xcrys}
Kokalj A 2003
{\it Comp. Mater. Sci.} {\bf 28} 155 Code available from http://www.xcrysden.org


\end{thebibliography}
\end{document}